# Electrical conductivity enhancement of epitaxially grown TiN thin films

Yeong Gwang Khim[1,2,†], Beomjin Park[1,2,†], Jin Eun Heo,[3,†] Young Hun Khim[1], Young Rok Khim[1], Minseon Gu[1], Tae Gyu Rhee[1,2], Seo Hyoung Chang[3], Moonsup Han[1,*], and Young Jun Chang,[1,2,*]

[1]Department of Physics, University of Seoul, Seoul 02504, Republic of Korea,

[2]Department of Smart Cities, University of Seoul, Seoul 02504, Republic of Korea,

[3]Department of Physics, Chung-Ang University, Seoul, 06974 Republic of Korea,

**Abstract**

Titanium nitride (TiN) presents superior electrical conductivity with mechanical and chemical stability and compatibility with the semiconductor fabrication process. Here, we fabricated epitaxial and polycrystalline TiN (111) thin films on MgO (111), sapphire (001), and mica substrates at 640°C and room temperature by using a DC sputtering, respectively. The epitaxial films show less amount of surface oxidation than the polycrystalline ones grown at room temperature. The epitaxial films show drastically reduced resistivity (~30 micro-ohm-cm), much smaller than the polycrystalline films. Temperature-dependent resistivity measurements show a nearly monotonic temperature slope down to low temperature. These results demonstrate that high temperature growth of TiN thin films leads to significant enhancement of electrical conductivity, promising for durable and scalable electrode applications.

* Corresponding author's e-mails:

Moonsup Han    mhan@uos.ac.kr,

Young Jun Chang    yjchang@uos.ac.kr

[†]Y. G. Khim, B. Park, and J. E. Heo contributed equally to this work.

# 1. Introduction

Metallic thin films have played essential roles for electrodes and channels in metal-oxide-semiconductor field effect transistors,[1,2] metal-insulator-metal capacitors,[3] neuromorphic devices,[4] solar cells,[5] and electrochemical devices.[6] Scaling semiconductor devices with a length scale of a few nanometers request advancement of metallic compounds or search for novel material candidates.[7,8] The candidates include elemental metals (Cu, Ni),[2,9] metal nitrides (TiN, VN, NbN),[10–12] metal-oxides ($SrRuO_3$, $LaNiO_3$),[13–17] metal-chalcogenides,[18–21] and MXenes.[7,22] The reduced dimensions of metal films result in increased resistance, the so-called "size effect". Therefore, thin metal films with low resistivity are needed for practical application in advanced semiconductor devices.

Titanium nitride (TiN) is a widely used electrode material with exciting properties, such as transparent metallicity,[10] plasmonic property,[23] and superconductivity.[24] It's been widely studied primarily due to high-temperature stability, atomistic diffusion barrier, and semiconductor compatibility. Many attempts have been made to modify the electrical conductivity of TiN layers with several different approaches, such as chemical substitution,[25] partial oxidation,[5] or epitaxial growth.[10,23,24] It is increasingly essential to control the interfaces between the TiN layer and the adjacent dielectric layer since the interface's crystal orientations or oxidation states influence the device's performance. Significantly, the first principles simulations predicted that the TiN (111) surface orientation plays a critical role in determining the enormous value of work function and the most oxidation resistance among other surface orientations of TiN.[26] Although several previous studies on tuning the TiN electrode layers, it is still necessary to examine the influence of surface orientation and crystalline quality for improving electrical conductivity and interface characteristics. For obtaining TiN thin films, many different growth methods, such as sputtering,[24] molecular beam epitaxy,[23] pulsed laser deposition,[27] and atomic layer deposition,[28] are employed.

Here, we present a systematic study of TiN thin films on MgO (111), Al$_2$O$_3$ (001), and mica substrates at both 640°C and room temperature (25°C). High-resolution X-ray diffraction (HRXRD) measurements show distinct crystalline quality between the growth temperatures. X-ray photoemission spectroscopy (XPS) data also show changes in surface oxidation states. Temperature-dependent resistivity measurements finally demonstrate the impact of crystalline quality on electrical resistivity.

## 2. Experiment

TiN thin films were directly grown on MgO (111), Al$_2$O$_3$ (001), and mica substrates by using a DC sputtering (Korea Vacuum Tech., Korea) using a 2-in. titanium target (99.999%) and pure nitrogen gas (99.999%). The base pressure of the deposition chamber was $6 \times 10^{-8}$ Torr. The deposition pressure was maintained at 2.5 mTorr with Ar (12 sccm) and N$_2$ (1.5 sccm) gases. To obtain uniform deposition of the films, the sample holder was rotated at 7 rpm during growth. During growth, the substrate temperature was held at 640 °C and 25 °C, and the DC power was set at 300 W. To obtain uniform deposition, the three kinds of substrates were placed on the sample holder simultaneously.

High-resolution X-ray diffraction (HRXRD, Bruker Discover D8) measurements were used for analyzing the crystal phases. Both theta-2theta and grazing-incidence measurements were carried out for epitaxial and polycrystalline films, respectively. X-ray photoemission spectroscopy (XPS) was measured with a monochromatic Al K$\alpha$ source (Thermo Fisher Scientific, Nexsa) without any further surface preparation. The electrical resistivity was obtained as a function of temperature using a closed-cycle cryostat (Seongwoo Instruments, 4K closed-cycle refrigerator). Measurements were carried out in the van der Pauw geometry with square samples (5 mm × 5 mm) and indium Ohmic contacts in the sample corners. The four-terminal resistance was measured using a current source (6220, Keithley Inc.) and a nanovoltmeter (2182, Keithley Inc).

## 3. Results and discussion

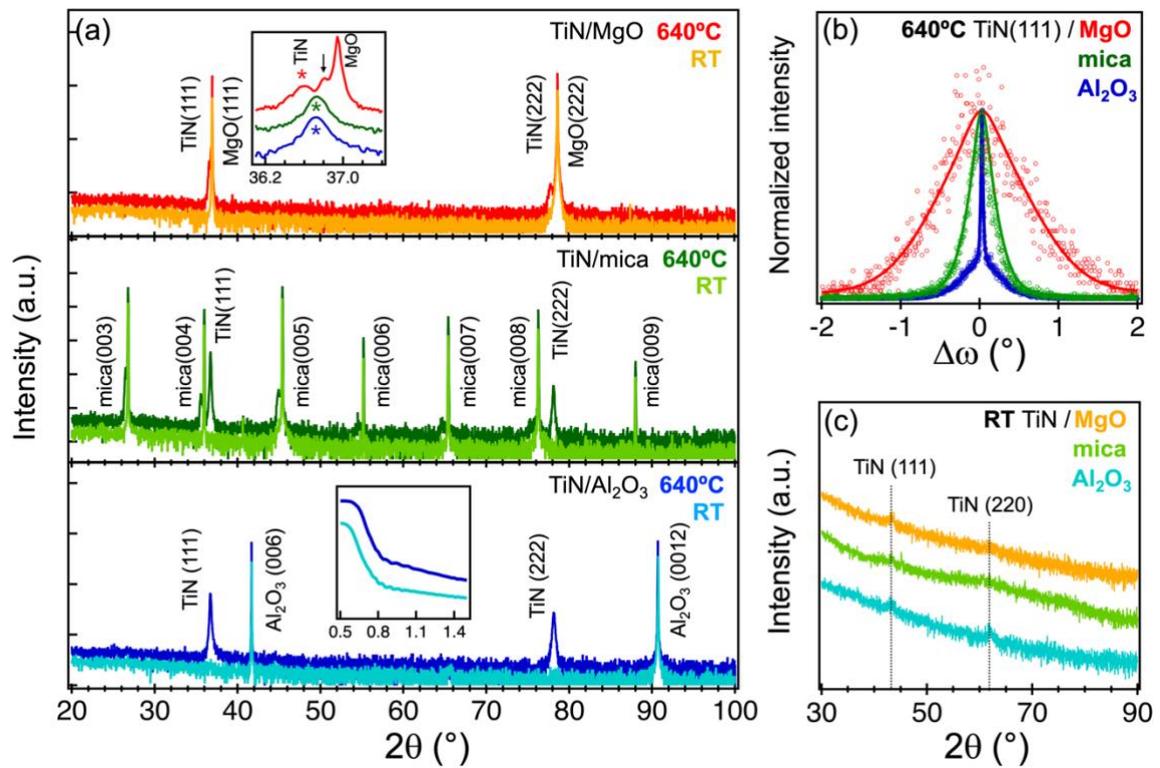

Figure 1. Crystal structure analysis of TiN thin films. (a) Theta-2theta measurements of the films grown at both 640°C and 25°C. The inset compares the TiN (111) peak positions. There are small thickness fringes near the TiN (111) peak. (b) Rocking curves of TiN (111) peaks for the epitaxial films grown at 640°C. (c) Grazing-incidence XRD curves of polycrystalline TiN films grown at room temperature.

In Figure 1, we examined the crystalline states of the TiN thin films from the HRXRD measurements. When grown at 640°C, all TiN films show both (111) and (222) peaks on MgO (111), Al$_2$O$_3$ (001), and mica substrates. The inset of Fig. 1(a) shows that the TiN (111) peaks (indicated by asterisks) locate at 36.6° - 36.7°, similar to the lattice parameter of previous TiN thin films (36.8°).[10,23] We note that the small peak (indicated by an arrow in the inset) between the TiN (111) and MgO (111) is due to thickness oscillation from the smooth film surface and bottom interface. Figure 1(b) shows the normalized rocking curves of the TiN (111) peaks. The different full-width half maximum (FWHM) peaks indicate that the TiN film is better aligned on the Al$_2$O$_3$ substrates (0.04°) than on the MgO (1.29°) and mica (0.4°). The narrow width of TiN/Al$_2$O$_3$ (001) is smaller than those of previously reported TiN/Al$_2$O$_3$ (0.22°).[23] It is also interesting that TiN film is also well (111)-oriented on the flexible mica substrate.[24] Considering lattice parameter differences between TiN and the substrates, MgO (0.684%), Al$_2$O$_3$ (6.43%), mica (13.3%), we speculate that TiN films may grow on both Al$_2$O$_3$ and mica substrates with nearly strain-relaxed state and show their lattice parameters slightly closer to the bulk value compared to the lattice of TiN on MgO. On the MgO substrates, however, we note that TiN film grows with compressive strain state. Therefore, the broad rocking curve of TiN/MgO (111) indicates that large fraction of TiN films is misaligned from the (111) orientation due to strain distribution inside the films.

On the other hand, when grown at room temperature, the TiN films show no diffraction peaks in the theta-2theta geometry. Instead, grazing incidence (GI)-XRD data (Fig. 1(c)) shows small peaks corresponding to the TiN phases. These minor peaks indicate that the room temperature sputtering gives rise to the formation of polycrystalline TiN phases. We note that the TiN/mica shows weaker peaks compared to the other two samples grown at room temperature. Therefore, we prepared epitaxial (111)-oriented and polycrystalline TiN thin films by simply tuning growth temperature. In addition, we estimated the film thicknesses to be ~80 nm and ~120 nm for the epitaxial and polycrystalline films from the x-ray reflectivity measurements (not shown), respectively.

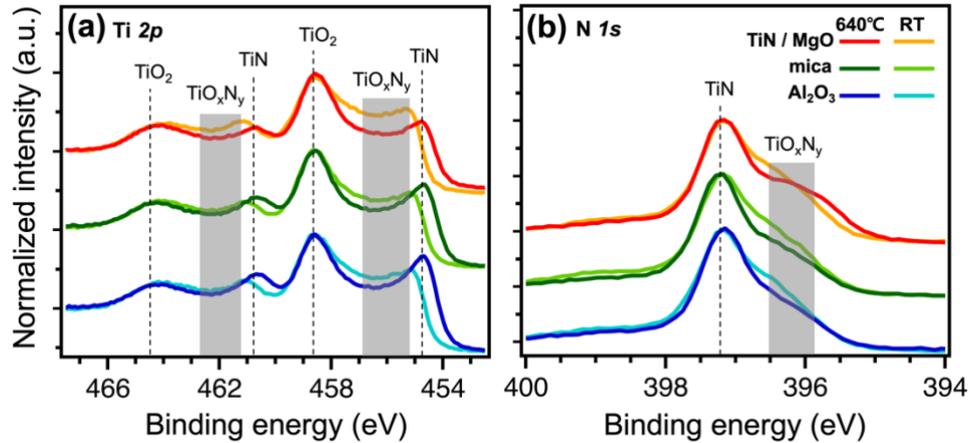

Figure 2. XPS analysis of TiN thin films with (a) Ti *2p* and (b) N *1s* states. TiN and $TiO_2$ phases show well-defined peaks, whereas intermediate $TiO_xN_y$ phases show broad features.

To analyze the surface stoichiometry, we performed XPS measurements (Fig. 2). We first note that all the TiN films show some surface oxidation due to air exposure. Figure 2(a) shows Ti 2p peaks comprised of TiN, $TiO_2$, and $TiO_xN_y$ bondings. Notably, the epitaxial films show well-separated peaks at 455 eV, corresponding to the TiN bonding. However, the polycrystalline films show suppression of TiN peak along with enhancement of $TiO_xN_y$ contributions. Due to variation of oxygen and nitrogen stoichiometry, the $TiO_xN_y$ peak is spread over a wide energy range, while the $TiO_2$ peak is well defined for all the samples.

Figure 2(b) shows N 1s peaks constituted with TiN and $TiO_xN_y$ bondings. The nitrogen peaks correspond to the nitrogen bonded to the N-Ti-N or O-Ti-N form. All films show a strong TiN peak, while the amount of weak $TiO_xN_y$ peak depends on the growth conditions. The epitaxial films show smaller oxynitride peaks for all substrates. Ti 2p and N 1s peaks indicate that the epitaxial films experience weaker oxidation at their surface region. As we discussed earlier about the XRD data, we note that significant number of tilted domains may also be present at the surface region of TiN/MgO, at which further oxidations are proceeded afterward in ambient exposure. This may explain the weakened TiN peak intensity at 455 eV (Fig. 2(a)) and the enhanced $TiO_xN_y$ peak at ~396 eV (Fig. 2(b)).

It is worth noting that TiN (111) has the highest work function compared to other crystal orientations making TiN (111) the most stable surface orientation as theoretically predicted.[26] Therefore, polycrystalline surfaces may experience more surface oxidation due to mixed surface crystal terminations and grain boundaries. Such differences of surface oxidation states and work function at the metal surfaces may influence metal-dielectrics interface properties.[29]

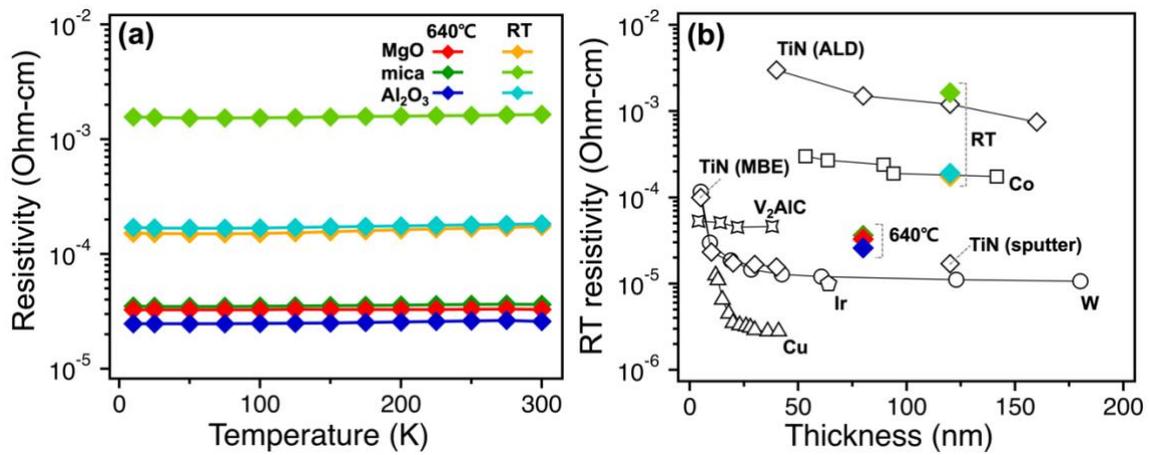

Figure 3. Electrical resistivity of TiN thin films. (a) Temperature-dependent resistivity of TiN thin films grown on different substrates at 640°C & RT(25°C). (b) Comparison of RT resistivity values for variolous electrode materials as a function of film thicknesses. (Cu,[30] Ir,[31] W,[32] Co,[33] $V_2AlC$,[7] TiN (MBE),[10] TiN (Sputter),[24] TiN (ALD)[34])

We then analyzed temperature-dependent electrical resistivity for the different TiN films. At first glance, all samples show monotonic temperature dependence with a small positive gradient, implying metallic character with some amount of scattering centers. Another feature is that the epitaxial films show much smaller resistivity (25 ~ 35 micro-ohm-cm) than the polycrystalline ones. The polycrystalline films show a considerable variation of resistivity depending on the substrates, MgO (150 micro-ohm-cm), $Al_2O_3$ (200 micro-ohm-cm), and mica (1500 micro-ohm-cm). It is reminded that the TiN/mica sample shows slightly weaker peak intensities in the GI-XRD data compared to the other two substrates. We speculate that the TiN/mica grown at RT might have either smaller volume of crystalline grains or smaller size distribution of grains, which further increases the electrical resistivity. Therefore, these results indicate that the resistivity of TiN films largely depends on the growth conditions, crystallinity, and substrates.

We compared such a large variation of electrical properties of TiN films with other metallic electrode materials as a function of film thicknesses, as shown in Fig. 3(b). The elemental metal films, such as Cu, Ir, and W, show very low resistivities over a wide thickness range.[30–32] But as the films become thinner than 20 nm, the resistivity sharply increases due to the size effect. On the other hand, MXene $V_2AlC$ thin films maintain its resistivity value low, even down to 5 nm.[7] Co thin films show relatively large value even for large thicknesses.[33] TiN thin films show a large variation of resistivity values depending on the growth methods, i.e., MBE (15 micro-ohm-cm for 40 nm), sputter (17 micro-ohm-cm for 120 nm), and ALD (750 ~ 3000 micro-ohm-cm).[10,24,34] Compared to the other metal thin films, the large resistivity gap between the epitaxial and polycrystalline TiN films implies that the performances of many electrical devices may be improved or modified by modifying the electrical properties of their electrode materials.[35] Finally, such significant variation of resistivity with crystallinity control in metal films could be helpful for a wide range of applications, such as wiring for semiconductor devices,[1,3,36] neuromorphic switching devices,[37,38] electrochemical electrodes,[39] solar cell electrodes,[5] and superconducting quantum electronics.[12,24]

## 4. Conclusion

TiN thin films were grown on different substrates by DC sputtering. As the growth temperature was elevated to 640°C, the electrical resistivity was drastically reduced compared to their room temperature counterparts. At 640°C, TiN thin films were epitaxially grown with the (111) orientation on MgO (111), $Al_2O_3$ (001), and mica substrates. The TiN (111) thin film on $Al_2O_3$ shows the narrowest rocking curve peak. TiN thin films were grown with polycrystalline states at room temperature regardless of substrates. The epitaxial (111)-oriented TiN films experience surface oxidation, but less oxidation occurs compared to the polycrystalline counterparts. Our observation suggests an alternative approach to tuning the electrical and crystalline properties of TiN thin films for various electrical conductor applications.

**Acknowledgements**

This research has been supported by the Ministry of Trade, Industry and Energy and the KSRC (20010569), NRF (2020R1A2C200373211, 2020R1C1C1012424), MOLIT as [Innovative Talent Education Program for Smart City]. This work was supported by the 2020 Research Fund of the University of Seoul for M.Han.

## REFERENCES

1. Lima, L. P. B., Moreira, M. A., Diniz, J. A. & Doi, I. Titanium nitride as promising gate

electrode for MOS technology. *Phys. Status Solidi C* **9**, 1427–1430 (2012).

2. Kim, S. J. *et al.* Flat-surface-assisted and self-regulated oxidation resistance of Cu(111). *Nature* **603**, 434–438 (2022).

3. Jung, M., Gaddam, V. & Jeon, S. A review on morphotropic phase boundary in fluorite-structure hafnia towards DRAM technology. *Nano Converg.* **9**, 44 (2022).

4. Tian, C., Wei, L., Li, Y. & Jiang, J. Recent progress on two-dimensional neuromorphic devices and artificial neural network. *Curr. Appl. Phys.* **31**, 182–198 (2021).

5. Peng, J. *et al.* Centimetre-scale perovskite solar cells with fill factors of more than 86 per cent. *Nature* **601**, 573–578 (2022).

6. Wang, S., Lu, A. & Zhong, C.-J. Hydrogen production from water electrolysis: role of catalysts. *Nano Converg.* **8**, 4 (2021).

7. Yoo, J. E. *et al.* MAX-Phase Films Overcome Scaling Limitations to the Resistivity of Metal Thin Films. *ACS Appl. Mater. Interfaces* **13**, 61809–61817 (2021).

8. Kim, J., Lee, J.-K., Chae, B., Ahn, J. & Lee, S. Near-field infrared nanoscopic study of EUV- and e-beam-exposed hydrogen silsesquioxane photoresist. *Nano Converg.* **9**, 53 (2022).

9. Blackburn, J. M., Long, D. P., Cabañas, A. & Watkins, J. J. Deposition of Conformal Copper and Nickel Films from Supercritical Carbon Dioxide. *Science* **294**, 141–145 (2001).

10. Ho, I. H. *et al.* Ultrathin TiN Epitaxial Films as Transparent Conductive Electrodes. *ACS Appl. Mater. Interfaces* **14**, 16839–16845 (2022).

11. Rai, S., Prajapati, A. K. & Yadawa, P. K. Effect of temperature on elastic, mechanical and thermophysical properties of VNx (0.76 ≤ x ≤ 1.00) epitaxial layers. *J. Korean Phys. Soc.* (2022) doi:10.1007/s40042-022-00643-3.

12. Yan, R. *et al.* GaN/NbN epitaxial semiconductor/superconductor heterostructures. *Nature* **555**, 183–189 (2018).

13. Lee, K. J. *et al.* Structural properties of ferroelectric heterostructures using coherent bragg rod analysis. *Curr. Appl. Phys.* **20**, 505–509 (2020).

14. Kwak, Y. M. *et al.* Magnetoresistance of epitaxial SrRuO3 thin films on a flexible CoFe2O4-buffered mica substrate. *Curr. Appl. Phys.* **34**, 71–75 (2022).

15. Li, J. *et al.* Effect of annealing temperature on resistive switching behavior of Al/ La0.7Sr0.3MnO3 /LaNiO3 devices. *Curr. Appl. Phys.* (2022) doi:10.1016/j.cap.2022.11.013.

16. Kim, H. J. *et al.* Chemical and structural analysis of low-temperature excimer-laser annealing in indium-tin oxide sol-gel films. *Curr. Appl. Phys.* **19**, 168–173 (2019).

17. Sohn, B. & Kim, C. Evolution of electronic band reconstruction in thickness-controlled perovskite SrRuO$_3$ thin films. *J. Korean Phys. Soc.* (2022) doi:10.1007/s40042-022-00633-5.

18. Lam, N. H. *et al.* Controlling Spin–Orbit Coupling to Tailor Type-II Dirac Bands. *ACS Nano* **16**, 11227–11233 (2022).

19. Duvjir, G. *et al.* Emergence of a Metal–Insulator Transition and High-Temperature Charge-Density Waves in VSe2 at the Monolayer Limit. *Nano Lett.* **18**, 5432–5438 (2018).

20. Duvjir, G. *et al.* Fine structure of the charge density wave in bulk VTe2. *APL Mater.* **10**, 111102 (2022).

21. Ranjan, P. *et al.* 2D materials: increscent quantum flatland with immense potential for applications. *Nano Converg.* **9**, 26 (2022).

22. Iqbal, A., Hong, J., Ko, T. Y. & Koo, C. M. Improving oxidation stability of 2D MXenes: synthesis, storage media, and conditions. *Nano Converg.* **8**, 9 (2021).

23. Guo, W.-P. *et al.* Titanium Nitride Epitaxial Films as a Plasmonic Material Platform: Alternative to Gold. *ACS Photonics* **6**, 1848–1854 (2019).

24. Zhang, R. *et al.* Wafer-Scale Epitaxy of Flexible Nitride Films with Superior Plasmonic and Superconducting Performance. *ACS Appl. Mater. Interfaces* **13**, 60182–60191 (2021).

25. Lou, M., Wang, L., Chen, L., Xu, K. & Chang, K. Role of refractory metal elements addition on the early oxidation behavior of TiN coatings. *Mater. Lett.* **331**, 133406 (2023).

26. A. Calzolari & A. Catellani. Controlling the TiN Electrode Work Function at the Atomistic Level: A First Principles Investigation. *IEEE Access* **8**, 156308–156313 (2020).

27. Torgovkin, A. *et al.* High quality superconducting titanium nitride thin film growth using infrared pulsed laser deposition. *Supercond. Sci. Technol.* **31**, 055017 (2018).

28. Jung, Y. W. *et al.* Study on TiN film growth mechanism using spectroscopic ellipsometry. *J. Korean Phys. Soc.* **80**, 185–189 (2022).

29. Seo, J., Cho, S. W., Ahn, H.-W., Cheong, B. & Lee, S. A study on the interface between an amorphous chalcogenide and the electrode: Effect of the electrode on the characteristics of the Ovonic Threshold Switch (OTS). *J. Alloys Compd.* **691**, 880–883 (2017).

30. Lim, J.-W., Mimura, K. & Isshiki, M. Thickness dependence of resistivity for Cu films deposited by ion beam deposition. *Appl. Surf. Sci.* **217**, 95–99 (2003).

31. Hämäläinen, J. *et al.* Atomic Layer Deposition of Iridium Thin Films by Consecutive Oxidation and Reduction Steps. *Chem. Mater.* **21**, 4868–4872 (2009).

32. Choi, D. *et al.* Phase, grain structure, stress, and resistivity of sputter-deposited tungsten films. *J. Vac. Sci. Technol. Vac. Surf. Films* **29**, 051512 (2011).

33. Pal, A. K., Chaudhuri, S. & Barua, A. K. The electrical resistivity and temperature coefficient of resistivity of cobalt films. *J. Phys. Appl. Phys.* **9**, 2261–2267 (1976).

34. Musschoot, J. *et al.* Atomic layer deposition of titanium nitride from TDMAT precursor.

*Microelectron. Eng.* **86**, 72–77 (2009).

35. Sun, C. *et al.* Control the switching mode of Pt/HfO2/TiN RRAM devices by tuning the crystalline state of TiN electrode. *J. Alloys Compd.* **749**, 481–486 (2018).

36. Kim, Y. S., Chung, H., Kwon, S., Kim, J. & Jo, W. Grain boundary passivation via balancing feedback of hole barrier modulation in HfO2-x for nanoscale flexible electronics. *Nano Converg.* **9**, 43 (2022).

37. Kim, T., Kim, Y., Lee, I., Lee, D. & Sohn, H. Ovonic threshold switching in polycrystalline zinc telluride thin films deposited by RF sputtering. *Nanotechnology* **30**, 13LT01 (2019).

38. Jay Kim, M. *et al.* Epitaxial growth and optical band gap variation of ultrathin ZnTe films. *Mater. Lett.* **313**, 131725 (2022).

39. Pedanekar, R. S., Shaikh, S. K. & Rajpure, K. Y. Thin film photocatalysis for environmental remediation: A status review. *Curr. Appl. Phys.* **20**, 931–952 (2020).